# Water on The Moon, I. Historical Overview
Arlin Crotts (Columbia University)

By mid-19th century, astronomers strongly suspected that the Moon was largely dry and airless, based on the absence of any observable weather. [1] In 1892, William H. Pickering made a series of careful occultation measurements that allowed him to conclude that the lunar surface's atmospheric pressure was less than 1/4000th of Earth's. [2] Any number of strange ideas arose to contradict this, including Danish astronomer/mathematician Peter Andreas Hansen's hypothesis, that the Moon's center of mass is offset by its center of figure by 59 kilometers, meaning that one or two scale-heights of atmosphere could hide on the far side of the Moon, where it might support water oceans and life. [3] Hans Hörbiger's 1894 *Welteislehre* ("World Ice") theory, that the Moon and much of the cosmos is composed of water ice, became the favored cosmology of leaders of Third Reich Germany. [4] Respectable scientists realized that significant amounts of water on the Moon's surface would rapidly sublime into the vacuum.

[1] "The absence of any bodies of water on the moon is placed beyond doubt, both by actual telescopic examination and by inference from absence of clouds. There are no streams, lakes or seas. An eminent astronomer has remarked that the heat of the surface exposed to the sun would occasion a transfer of any water the moon might contain to its dark side, and that there may be frosts in this part, and perhaps running water near the margin of the illuminated portion. But in such a case, would not clouds appear about the margin at times in telescopic views?" ("On the Volcanoes of the Moon" by James D. Dana, 1846, *American Journal of Science & Arts*, 11, 335.)

[2] "The Lunar Atmosphere and The Recent Occultation of Jupiter" by William H. Pickering, 1892, *Astronomy & Astrophysics, Carleton College Goodsell Observatory*, 11, 778.

[3] "Sur la figure de la lune" by Peter A. Hansen, 1856, *Memoirs of the Royal Astronomical Society*, 24, 29.

[4] *Glacial-Kosmogonie* by Hans Hörbiger, 1912, Leipzig: Voigtländer, 772 pp.

In the mid-20th century, the prospect of water on the Moon was adopted by a more competent and respected scientist, Harold Urey, recipient of the 1934 Nobel Prize in Chemistry for discovering deuterium (heavy isotope of hydrogen), and noted for contributions to uranium isotopic enrichment, cosmochemistry, meteoritics and isotopic analysis, and the idea that life might have arisen from the chemistry of Earth's primordial atmosphere.  In the early days of the space program, Urey was decisive in persuading NASA to scientifically explore the Moon. [5] He began writing about the Moon in 1952, in his book *The Planets*, arguing that to understand planetary origins we should study the Moon, the "Rosetta Stone of the Solar System" – a primitive world that might hold samples indicative of the early Earth.  He based this in part on the idea that the Moon and planets perhaps formed by accretion of cold material like dust.  In such a scenario he imagined water playing important roles.

[5] "Harold Urey & The Moon" by Homer E. Newell, 1972, *Earth, Moon & Planets*, 7, 1; "Nickel for Your Thoughts: Urey & the Origin of the Moon" by Stephen G. Brush, 1982, *Science*, 217, 891.

   Urey imagined many processes influencing the lunar surface.  By 1956 he was entertaining the effects of water on the Moon upon its surface features. [6] His thinking was influenced by new data, particularly pictures of the Moon starting with *Ranger 7* in 1964. [7] *Ranger 9* in 1965 returned close-up photos of the flooded crater Alphonsus revealing a complex of atypical features for the Moon, including rivers and channels, some appearing cut by flowing liquid. (See Figure 1.) Urey jumped to their interpretation: "Various lines of evidence indicate that the material of the maria floors and especially of the Alphonsus floor consist of fragmented material to a very considerable depth, with substantial crevasses below the surface.  It is not possible to decide whether such crevasses are the result of lava flows or the evaporation of massive amounts of water from beneath the surface.  It is, however, the author's opinion that the water interpretation is the more likely of the two."  He summarized: "They (lunar maria) may have been subjected to water at some time in their history, but the evidence of pictures alone is not sufficient to make a firm decision in regard to these conclusions." [8]

[6] "The Origin and Significance of the Moon's Surface" by Harold C. Urey, 1956, *Vistas in Astronomy*, 2, 1667.  See also "The Moon's Surface Features" 1956, *Observatory*, 76, 232 which

among other ideas suggests exploding an atomic bomb on the Moon's surface and collecting the resulting lunar meteorites reaching Earth!

[7] Urey had immediate access to the data, being one of the few Ranger project co-experimeters: "But if water were present on the moon, one may ask how much and for how long. Since river valleys or stream structures of any kind are not present on the moon, it seems certain that the amount was small and the time was short. Small effects of this kind could have been destroyed by the erosion processes shown to be present by the Ranger 7 pictures. Could it be that the comparatively smooth floors of the maria are the beds of ancient temporary lakes? Their smooth structure has led most students of the subject to assume that the maria are lava flows, and anyone not subscribing to this view is compelled to try to devise other explanations for this smoothness. The Ranger 7 pictures have made many people, including me, think seriously that Mare Cognitum consists of fragmented material rather than lava flow material. We must account for the crater Wargentin, which is full of smooth material to the brim. Could it be water or ice covered with some layer of dust and could it have become filled with water by temporary rains, and are its walls impervious to water while those of other craters are not? It has always seemed odd to me that the moon could produce hot lavas to fill Wargentin and at the same time be sufficiently rigid to support differences of 10 km in elevation of the lunar surface. Kopal and Gold have proposed that water has diffused from the lunar interior to fill the maria basins, and they compare this to water coming from the interior of the earth. However, water probably comes from the earth's interior through its numerous volcanoes and lava flows which have covered the original surface of the earth to a mean depth of some 15 km. Craig finds that terrestrial hot springs consist mostly or entirely of meteoric and not juvenile water. It is difficult to believe that diffusion of water through rocky material either of the earth or moon would supply more than very limited quantities of water to their surfaces. Also the estimates of the amounts of water that have come to the earth's surface during geological time rest on very uncertain evidence. But these suggestions of Kopal and Gold have stimulated me to consider the possibility that contamination of the moon with water from the earth was larger than I intended to suggest previously. Only the Surveyor and Apollo missions to the moon can answer the questions raised in this way." ("Meteorites and the Moon" by Harold C. Urey, 1965, *Science*, 147, 1262.)

[8] "Study of the Ranger Pictures of the Moon" by Harold C. Urey. 1967, *Proceedings of the Royal Society of London, Series A*, 296, 418.

Urey was not sole proponent of a watery Moon. A topic common to several authors was the similar look of lunar sinuous rilles and terrestrial meandering rivers, with their oxbow loops (Figure 2). These need not form by water but by any sufficiently non-viscous, turbulent fluid. When a river rounds a bend, water is forced outward centrifugally so a higher water level rides in outer radii of the bend than inner ones. On the river bottom, however, water flows slowly due to drag from the riverbed. That water feels the high pressure on the outside of the bend, so flows towards inner radii, taking silt with it. Thus the riverbank erodes from outer radii and the oxbow grows. This also occurs for lava if the ratio of viscous force acting on the lava divided by the lava's inertia of motion (called the *Reynolds number*) is sufficiently similar to that ratio for terrestrial rivers. [9] Lunar basaltic lava is so less viscous than many on Earth that lunar lava flows and terrestrial meandering rivers share common turbulence conditions. Their resemblance arises from similar physical parameters, not the same fluid. Evidence for lunar water does not depend on observing landforms.

[9] If L is the size of the system being considered, the rate at which momentum density $\rho v$ (where $\rho$ is mass density and v the velocity) flows through area $L^2$ at velocity v is $\rho v^2 L^2$. Viscous force on an object at low velocity is proportional to v, size of the object L, with proportionality constant given by the viscosity $\mu$, so $F_{drag} = \mu vL$. The Reynolds number R is the ratio of momentum density to drag force: $R = \rho v^2 L^2 / \mu vL = \rho vL / \mu$, which is dimensionless. If $R \gtrsim 500$, the flow is turbulent, usually the case for water. Basaltic magma is at least about 10,000 times more viscous than water. Magma that is *rhyolitic* (alternatively *siliceous* – high in silica) can have $\mu$ values up to 10 million times water's. R will exceed 500 for basaltic lava only if the flow width L and mass flow rate (which determines v) are also large, but almost never does so for siliceous magma. ("Lunar Sinuous Rille Formation by Thermal Erosion: Conditions, Rates and Duration" by J.W. Head & L. Wilson, *Abstracts of Lunar & Planetary Science Conference*, 12, 427, and "The Formation of Eroded Depressions around the Sources of Lunar Sinuous Rilles: Theory" by J.W. Head & L. Wilson, *Abstracts of Lunar & Planetary Science Conference*, 11, 1260.)

Urey suffered sharp criticism for his openness to ideas of water on the Moon. It wore on him, so much that he confessed (in a letter to the prestigious journal *Nature*) that some thought him under the influence of more intoxicating liquids. [10] The most substantial attack came from NASA planetary scientist John O'Keefe (who suffered his own controversy regarding the lunar origin of glassy globules called *tektites*, now thought to ejected from large terrestrial impacts). O'Keefe showed how an ice layer under the lunar surface would distort and flow like a glacier if it exceeded a kilometer thickness, in contradiction to craters two kilometers deep absent signs of their walls flowing onto their floors. [11] The criticism is not definitive: a thinner layer would not deform but might conceivably produce effects Urey claimed in Ranger photos. Urey grew haggard defending himself and authors of similar ideas. [12]

[10] "The possibility that water has existed on the Moon for varying lengths of time, both in liquid and in solid form, and both beneath the surface and on the surface, has been widely

discussed during the past 10 years. The subject has been discussed repeatedly at scientific meetings and has been received mostly with great skepticism. Evidence supporting this view has recently become quite overwhelming and, in fact, no communication seems necessary to point out the evidence from the *Orbiter 4* and *5* pictures. Because many people are not aware of this evidence and suggest that the effects are caused by other liquids, that is, lava, dust-gas or possibly even vodka, a brief discussion of the evidence may be in order." ("Water on the Moon" by Harold C. Urey, 1967, *Nature*, 216, 1094)

[11] "Water on the Moon and a New Nondimensional Number" by John A. O'Keefe, 1969, *Science*, 163, 669.

[12] Urey retorted to O'Keefe: "All right, attack if you wish to. This is, so far as I recall, my suggestion, not that of my good friend, T. Gold. Possibly Lingenfelder et al. considered some modification of this idea. I am not at all convinced that Gold's mechanism may not contribute to the problem to some extent." ("Water on the Moon" by Harold C. Urey, 1969, *Science*, 164, 1088.) Lingenfelter et al. postulated that a 1-kilometer thick ice layer on the Moon, shielded from sublimation into the vacuum by a 100-meter overburden of regolith, could melt when impacted by a meteorite and flow underground to form rilles as in Figure 2. ("Lunar Livers" by Richard E. Lingenfelter, Stanton J. Peale & Gerald Schubert, 1968, *Science*, 161, 266). Gold detailed similar icesheet/overburden geometry to explain odd-looking lunar craters which seemed to contain features similar to *pingos*: ice-mounds caused by water forced out of permafrost. ("The Moon's Surface" by T. Gold, 1966, in *The Nature of the Lunar Surface*, edited by Wilmot N. Hess, Donald H. Menzel & John O'Keefe, Baltimore: Johns Hopkins Press, p. 107.)

This changed with the first samples returned from the Moon, with almost no signs of hydrated minerals. The only layered lunar rocks result from layered lava. Hadley Rille, visited by *Apollo 15*, was obviously made by flowing lava, not water. Urey relented, abandoning any idea of lunar water. [13] Eventually, Urey and O'Keefe published scientific papers together but none about lunar hydration. Urey died in 1981, long before minds began to change regarding water on the Moon.

[13] From discussion after an address by Harold Urey in Philadelphia on April 24, 1970:
Questioner: "What is the origin of the crooked rilles? Possibly water?"
Urey: "The crooked rilles on the Moon looked to me, for a long time, as though they were due to liquid water. Since water cannot be found in the rocks, these rilles must have been produced by melted silicates of some kind flowing across the surface of the Moon. I am immensely surprised that such materials as this would flow for 250 kilometers across the surface of the Moon without freezing, and at the end, just disappear somewhere. On the other hand, other people suggest that the rilles are due to lava flows from the interior of the Moon. In fact, I think we must be rather skeptical in regard to all sources of water on the Moon, unless we get other evidence pointing in a different direction; water on the Moon seems not to have been present in any important amounts at any time in the past. We may have to revise that opinion again in the future, but that certainly is my opinion at the present time." ("A Review of the Structure of the Moon" by Harold C. Urey, 1971, *Proceedings of the American Philosophical Society*, 115, 67.) Urey wrote more papers about the Moon, but never again about water. Even Tommy Gold at this April 1970 symposium ceased mention of water, but persisted in his theory of the lunar surface dominated by dust transported by electrostatic force. ("The Nature of the Lunar Surface: Recent Evidence" by Thomas Gold, 1971, *Proceedings of the American Philosophical Society*, 115, 74)

The evidence from Apollo against water on the Moon was varied and manifest, and there was little in favor. Even though astronauts left instruments on the Moon detecting a substantial lunar atmosphere, most of its mass was the inert gas argon, formed in the decay of radioactive potassium. [14] The amount of water was less than 1% of this, barely detectable at a density of 600 molecules per cubic centimeter. [15] Some unknown gas had been dissolved in the lava that flowed onto the lunar maria and expanded into foamy *vesicular basalt* as the lava reached the surface and cooled, with the gas expanding under lower pressure. (See Figure 3.) If these vesicles were full of steam, the mineral that should result is *amphibole*, which is essentially like the mineral pyroxene, common on the Moon, but hydrated. The absence of amphibole in vesicular basalts indicated the gas within the lava had consisted almost entirely of something other than water. [16]

[14] $^{40}$K decays with a half-life of 1.277 billion years, either becoming $^{40}$Ca by emitting an electron (β particle) in 89.28% of decays, or by capturing an electron out of its atomic orbital to become $^{40}$Ar in 10.72% of decays (unless the $^{40}$K is ionized). Thus after 1.277 billion years, from a sample of $^{40}$K atoms, 50% will still be $^{40}$K, 44.64% will be $^{40}$Ca, and 5.36% will be $^{40}$Ar, an inert gas at temperatures above 87.3 K at 1 atmosphere pressure, and above 150.87 K at pressures under 50 atmospheres).

[15] "Molecular gas species in the lunar atmosphere" by J.H. Hoffman & R.R. Hodges, Jr., 1975, *The Moon*, 14, 159. The molecular water number density at sunrise (the peak signal) is 600 ± 300 cm$^{-3}$ (1σ) versus the $^{40}$Ar signal of 30000 cm$^{-3}$ at sunrise, which drops to under 1000 cm$^{-3}$ at night.

[16] "H$_2$O in lunar processes: The stability of hydrous phases in lunar samples 10058 & 12013" by R.W. Charles, D.A. Hewitt & D.R. Wones, 1971, *Proceedings of 2nd Lunar Science Conference*, 1, 645.

Arguments developed that lunar rocks contained essentially no water at all; the water content of the Moon could be limited to a level that was in the range of *parts per billion* or even less. There were exceptions; in fact many samples contained water at the level of 250 to 500 parts per *million* (by weight). [17] The H$_2$O in samples was identical to "Pasadena water vapor" in a way noted by the Caltech authors of one analysis: the ratios of the different types, or isotopes, of oxygen and hydrogen were the same as on Earth. [18] On Earth, the nuclei of most hydrogen atoms consist of just a proton, but 0.015% of hydrogen nuclei have a neutron attached to this proton (hence are deuterium: $^2$H), and 0.2% of oxygen has an extra two neutrons ($^{18}$O, versus the usual isotope $^{16}$O). On the Moon, oxygen isotopes are essentially in the same ratio as on Earth, so the argument becomes one mainly of deuterium. [19] (Remember: 45% of the lunar soil is composed of oxygen.) The deuterium ratio in the solar wind striking the Moon is not well known, but much smaller than on Earth. Many lunar samples have a tiny deuterium fraction, so their hydrogen is assumed to derive from solar wind. If they show a high hydrogen abundance, they also tend to have a larger deuterium fraction, close to the terrestrial value, which is interpreted as contamination. Even minerals that seem to evidence aqueous processing e.g., rust, show the same isotope ratios, so were suspected to result from terrestrial water. [20]

[17] "Deuterium, hydrogen and water content of lunar material" by L. Merlivat, M. Lelu, G. Nief & E. Roth, 1974, *Geochimica et Cosmochimica Acta*, 2, 1885.

[18] "$^{18}$O/$^{16}$O, $^{30}$Si/$^{28}$Si, $^{13}$C/$^{12}$C & D/H studies of *Apollo 14 & 15* samples" by S. Epstein & H.P. Taylor, Jr., 1972, *Proceedings of 3rd Lunar Science Conference*, 2, 1429. We discuss more recent measurements of D/H shortly, which tell a slightly different story.

[19] If one strips away increasingly deeper layers of regolith grains (by reacting them with fluorine gas), the $^{18}$O/$^{16}$O ratio is at first very high, then decreases to the terrestrial value as center of the grains are reached. ("$^{18}$O/$^{16}$O, $^{30}$Si/$^{28}$Si, $^{13}$C/$^{12}$C & D/H Studies of *Apollo 16* Lunar Samples' by S. Epstein & H.P. Taylor, Jr., 1973, *Abstracts of Lunar & Planetary Science Conference*, 4, 228)

[20] "D/H & $^{18}$O/$^{16}$O ratios of H$_2$O in the 'rusty' breccia 66095 & the origin of 'lunar water'" by S. Epstein & H.P. Taylor, Jr., 1974, *Proceedings of 5th Lunar Science Conference*, 2, 1839.

The problem with this interpretation is there is no way one could conclude water was lunar if it had the same oxygen and hydrogen isotopic ratios on Earth and the Moon, as is the case with so many chemical elements, and any lesser fraction of deuterium could be explained away by solar wind origin.  The hypothesis of a dry Moon lacks enough scientific falsifiability – a wide range of hydrous isotopic ratios could be explained in terms solar wind origin and terrestrial contamination regardless of whether the water originated in the Moon.

Was terrestrial contamination a problem?  Definitely, but one that has been exaggerated.  Dust covers essentially all surfaces on the Moon, and quickly extends to equipment (and people) from Earth unless carefully avoided.  This dust led to terrestrial contamination in a subtle way: dust fouled the air seals on some sample boxes intended to bring lunar dust and rocks back to Earth isolated from the air (and moisture). [21] Apollo Lunar Sample Return Containers, used to carry soil and rock home on all six landing missions, were aluminum boxes with triple seals protected by a cloth and Teflon cover removed just prior to closing.  Even so, some of the seals were compromised by the ubiquitous coating of hard-grained dust. [22] Samples were held in passive quarantine for six weeks along with returning astronauts to prevent lunar biological infection, and over the years samples have been contaminated in processing and analysis.  This issue grew to the common myth that all samples were corrupted with terrestrial atmosphere, [23] which is wrong.  Even now some Apollo samples almost certainly remain uncontaminated since being contained. [24] Unfortunately, one can also imagine mild contamination caused by the Lunar Module's exhaust, or by astronauts' backpacks, which vent water vapor in order to cool. [25]

[21] From Alan Bean after *Apollo 12*: "Closing of the sample return containers was not difficult and was similar to that experienced during one-sixth g simulations in an airplane.  The seal for the sample return container lid became coated with considerable dust when the documented samples were being loaded into the container.  Although the surface was then cleaned with a brush, the container did not maintain a good vacuum during the return to earth." – (*Apollo 12 Mission report*, Manned Space Center, Houston, March 1970, NASA MSC-01855, p. 9/19)

[22] From Lunar Curator of samples at NASA Johnson Space Center, Gary Lofgren: "Most of the Apollo Lunar Sample Return Containers (ALSRCs) sealed."  He lists Earth-received status of all 12 boxes; one of two ALSRCs on each of *Apollo 12*, *14*, *15* & *16* leaked; eight maintained under 0.00025 atmosphere.  ("Overview & Status of the Apollo Lunar Collection" by Gary E. Lofgren, 2009, *Lunar Exploration & Analysis Group* meeting, #2075; also *Catalog of Apollo Experiment Operations*, Thomas A. Sullivan, 1994 January, NASA Reference Publication 1317, p. 60)

[23] For instance: "Many also believe a lunar sample return will be necessary.  True, the Apollo astronauts brought back some 800 pounds of lunar rocks from six landing sites.  But the dust played a dirty trick: The gritty particles deteriorated the knife-edge indium seals of the bottles that were intended to isolate the rocks in a lunar-like vacuum.  Air has slowly leaked in over the past 35 years.  'Every sample brought back from the moon has been contaminated by Earth's air and humidity,' Olhoeft says.  The dust has acquired a patina of rust, and, as a result of bonding with terrestrial water and oxygen molecules, its chemical reactivity is long gone."  ("Stronger Than Dirt: Lunar explorers will have to battle an insidious enemy—dust" by Trudy E. Bell, 2006 September 1, *Air & Space Magazine*, http://www.airspacemag.com/space-exploration/dust.html)

[24] Special Environment Sample Containers (SESCs) excluded atmospheric contamination (with a stainless steel edge pressed into an indium seal mentioned in [23]).  About half of samples in the six SESCs are unused, with one container completely unopened.  Four other containers for regolith cores and other samples are also pristine.  ("Special, Unopened Lunar Samples: Another Way to Study Lunar Volatiles" by Gary E. Lofgren, 2011, *A Wet vs. Dry Moon: Exploring Volatile Reservoirs & Implications for the Evolution of the Moon & Future Exploration*, #6041)

[25] "…that's terrestrial water, which has a much, much larger amount of deuterium.  We kept on trying to get down to the least contaminated samples, but we couldn't find any lunar soil sample that didn't have a little tiny bit of deuterium in it, and we finally concluded that it was just contamination from the astronauts' backpacks, because the samples were carefully preserved on the way back.  And after they got to Houston, we don't think it was added, although you can never be sure, because it doesn't take much contamination to put a little bit of deuterium in there."  ("Hugh P. Taylor, 1932 – " interviewed by Shirley K. Cohen, 2002 June – July, Archives of California Institute of Technology, Pasadena, p. 83)

   Beyond isotopic ratios, there were further reasons to think the Moon contains little water.  For instance, lunar rocks are depleted of low melting-point elements such as lead, thallium, bismuth and indium (melting at 327, 304, 272 and 157ºC, respectively), to a level only about 1% of that in terrestrial basalts.  Since these would accrete last onto the Moon as it cooled, one might use these as a guide to water abundance (assuming that they all accrete from carbonaceous chondritic meteorites, for instance).  This implies water of only a few tenths part per million, equivalent to an ocean only 3.7 meters deep spread over the entire lunar surface. [26] Of course the result might be much different, and wetter, if the water arrived via comets.  With few exceptions, however, the Apollo samples seemed to contain much less water than terrestrial rocks, and the bulk of that was suspected to be terrestrial contamination. [27] Three decades passed before the issue was revisited, and changed dramatically.

[26] "Water on the Moon?" by Edward Anders, 1970, *Science*, 169, 1309.

[27] To put this is context, Earth's igneous minerals range from tens of ppm to over 5% water. Fresh basalts on mid-ocean ridges fall in the range 0.1 – 0.5% water, but those formed by hot mantle plumes range over several times wetter ("Recycled dehydrated lithosphere observed in plume-influenced mid-ocean-ridge basalts" by Jacqueline Eaby Dixon, Loretta Leist, Charles Langmuir & Jean-Guy Schilling, 2002, *Nature*, 420, 385). Earth's upper mantle is thought to have water in the range of 20 – 200 ppm ("Water in Anhydrous Minerals of the Upper Mantle: A Review of Data of Natural Samples and Their Significance" by Anne H. Peslier, 2007, *Workshop on Planetary Basalts*, #2003). "Dry as a bone" is not so dry: animal bone matrix is about 0.55% water! ("Water Content Measured by Proton-Deuteron Exchange NMR Predicts Bone Mineral Density & Mechanical Properties" by Maria A. Fernández-Seara, Suzanne L. Wehrli, Masaya Takahashi & Felix W. Wehrli, 2004, Journal of Bone & Mineral Research, 19, 289).

Of course, the Giant Impact lunar formation hypothesis, growing into favor in the 1970s and 1980s, explained why so little remained of volatile elements (lead, etc.) or compounds, like water. They boiled away into space. Thus, with hydrogen evaporated and the Moon's core largely lost inside Earth, only intermediate-mass elements remain abundant, starting with oxygen, which binds tightly to some heavier elements like silicon. The average temperature in pre-lunar material reached several thousand Kelvins, at which any mineral would melt and tend to release its dissolved gas. The Big Whack and Dry Moon were copacetic together.

Exceptions to the Dry Moon were seen even during the Apollo era. "Rusty Rock" 66095 discussed [20] is the most volatile-laden Apollo rock, an impact melt breccia typical of highlands rock. It has rust (often goethite: $FeO(OH)$), hence hydration, and schreibersite ($(Fe,Ni)_3P$), common in meteorites but otherwise largely absent from Earth. Since Rusty Rock seems to be the result of meteoritic impact, it was dismissed as hydrated by the meteorite. Most highlands rocks are breccias, however, and most of these contain both rust and schreibersite. Volatiles in highlands rock might be alien in origin, but they are not hard to find. [28]

[28] "Rusty Rock 66095 - A paradigm for volatile-element mobility in highland rocks" by Robert H. Hunter & Lawrence A. Taylor, 1982, *Lunar & Planetary Science Conference*, 12, 261, and "Rust and schreibersite in Apollo 16 highland rocks - Manifestations of volatile-element mobility" by Robert H. Hunter & Lawrence A. Taylor, 1982, *Lunar & Planetary Science Conference*, 12, 253.

Minerals are found occasionally in lunar rocks that usually indicate water's presence: hematite ($Fe_2O_3$), magnetite ($Fe_3O_4$) and goethite, and are more consistent with a low-temperature mix with hydrogen (a few hundred degrees Centigrade) than cometary impact. [29] Relics of comet impacts are also found; soil sample 61221 from *Apollo 16* contains $H_2O$, $CO_2$, $H_2$, $CH_4$, CN and perhaps CO, totaling 300 part per million. [30]

[29] "The origin and stability of lunar goethite, hematite and magnetite" by Richard J. Williams & Everett K. Gibson, 1972, Earth & Planetary Science Letters, 17, 84. See also "The Occurrence of Geothite in a Microbreccia from the Fra Mauro Formation" by S.O. Agrell, J.H. Scoon, J.V.P. Long & J.N. Coles, 1972, Abstracts of the Lunar & Planetary Science Conference, 3, 7.

[30] "Volatile-Rich Lunar Soil: Evidence of Possible Cometary Impact" by Everett K. Gibson, Jr. & Gary W. Moore, 1973, *Science*, 179, 69.

Not all evidence for lunar hydration is bound in rock and soil samples; sometimes it flies in from the vacuum. On *Apollo 14* astronauts Shepard and Mitchell installed SIDE (the Suprathermal Ion Detector Experiment) to detect accelerated, charged atoms in the lunar atmosphere and measure their mass and energy. Other versions of SIDE were set up by *Apollo 12* and later *Apollo 15*. On March 7, 1971, twenty-nine days after the *Apollo 14* Lunar Module *Antares* blasted off the lunar surface, another blast hit SIDE, millions of ions were detected, all with energies (about 49 electron volts) several times that sufficient to blow a typical atom apart, and with masses equal to 17 atomic mass units, the same as hydroxyl ions. [31] The SIDE at *Apollo 12*, 183 kilometers to the west also detected a blast of ions (but unfortunately was in a mode to measure the total number of ions, not their mass). The event was also detected by another experiment at *Apollo 14* as well. [32] In their 1973 paper the experimenters consider several artificial and extra-lunar sources and rule them all out except for natural outgassing from the lunar surface. Eighteen years later, however, two of the authors revisited the question and found that the most likely source of the water vapor was *Apollo 14*'s engine exhaust. [33] *Antares* had landed in the morning of the previous lunar day, and the March 7 event started about 30 hours after the Sun rose again over the site. Freeman and Hills argue that water vapor from *Apollo 14* sequestered itself into the lunar regolith over the previous lunation and was released by heating after sunrise.  This also was speculative, but it canceled out the original notion of natural hydroxyl outgassing, in most concerned people's minds.

[31] "Observations of Water Vapor Ions at the Lunar Surface" by J.W. Freeman, Jr., H.K. Hills, R.A. Lundeman & R.R. Vondrak, 1973, *Earth, Moon & Planets*, 8, 115.
    Another event on 1973 February 22 on LACE (Lunar Atmospheric Composition Experiment) at *Apollo 17*'s site showed un-ionized $CH_4$, $C_2H_6$ and $N_2$ (or CO), apparently releasing 10-500 kg of gas, 100-300 km from *Apollo 17*, a crowded neighborhood: *Luna 21* is in this annulus, with *Apollo 11*, *Surveyor 5*, *Ranger 6* and *8* within some 600 km.  Many contamination sources should be considered.  ("Summary of Conference: Interactions of Interplanetary Plasma with the Modern & Ancient Moon" by David R. Criswell & John W. Freeman, 1975, *The Moon*, 14, 3.  See p. 12.)

[32] The Charged Particle Lunar Environment Experiment is sensitive to ions and electrons between 60 eV and 50,000 eV in energy, coming from two directions (the zenith and 60° west of zenith).  The particles were detected in the low energy channel (60 – 300 eV per electronic charge).

[33] "The Apollo Lunar Surface Water Vapor Event Revisited" by J.W. Freeman, Jr. & H.K. Hills, 1991, *Geophysical Research Letters*, 18, 2109.  In 2009 with new evidence of lunar surface hydration water, Freeman said the new findings were consistent with their earlier results, which seems to contradict Freeman & Hills 1991. ("Water on the moon? Pfft. We saw that 40 years ago" by Eric Berger, 2009 September 24, *Houston Chronicle*: http://blog.chron.com/sciguy/2009/09/water-on-the-moon-pfft-we-saw-that-40-years-ago/)

By the early 1970s the Dry Moon was settled science, especially among lunar geologists in the United States.  Had that gone differently, the last mission of the Moon Race, by the Soviets, might have changed everything.  It discovered water *within* the Moon.  Instead it was ignored.

On August 9, 1976, *Luna 24* launched toward the Moon on a Proton rocket, and nine days later landed safely in the southern part of the unexplored Mare Crisium. [34] Within 24 hours, it deployed a drilling rig, extracted a core sample from two meters into the Moon, stowed it in its return capsule, and blasted off again with 170 grams of lunar soil.  Four days later it successfully re-entered the Earth's atmosphere over Siberia, and the core sample was taken to Moscow intact and uncontaminated (as far as we know).  It was the last lunar mission of the Soviet Union, and the last from Earth to soft-land on the Moon in the 20th century.

[34] *Luna 15* and *Luna 23* had both landed in Mare Crisium but failed.

What it brought back was very special. The core sample was found by scientists M. Akhmanova, B. Dement'ev, and M. Markov of the Vernadsky Institute of Geochemistry and Analytic Chemistry to contain about 0.1% water by mass, as seen in infrared absorption spectroscopy (at about 3 microns wavelength), at a detection level about 10 times above the threshold. The trend was for the water signal to increase looking deeper below the lunar surface. The original title of their paper in the February 1978 Russian language journal *Geokhimiia* translates to "Water in the regolith of Mare Crisium (Luna-24)?" and in the English language version of the journal "Possible Water in Luna 24 Regolith from the Sea of Crises"— but the abstract claims a detection of water fairly definitively. The authors point out that the sample shows no tendency to absorb water from the air, but they were not willing to stake their reputations on an absolute statement that terrestrial contamination was completely avoided. Nonetheless, they claim to have taken every possible precaution and stress that this result must be followed up. [35] The three Soviet lunar sample return missions (*Luna 16*, *20* and *24*) from 1970 to 1976 brought back a total of 327 grams of lunar soil. The six Apollo lunar landing missions in 1969 – 1972 returned 381,700 grams of rock and soil. Apollo won the samples race. No other author has ever cited the *Luna 24* work, as of this writing.

[35] "Water in the regolith of Mare Crisium (Luna 24)?" by M.V. Akhmanova, B.V. Dement'ev & M.N. Markov, 1978 February, *Geokhimiia*, 285, and "Possible Water in Luna 24 Regolith from the Sea of Crises" by M.V. Akhmanova, B.V. Dement'ev & M.N. Markov, 1978, *Geochemistry International*, 15, 166. I communicated with Boris Dement'ev in March 2010 about the integrity of the Luna 24 core sample, and he said that the sample might have absorbed water vapor in the laboratory, but their tests showed that it had little tendency to do so.

A generation passed without further work on lunar hydration. After Apollo the next lunar science mission was *Clementine* in January 1994. [36] Along with many distinctions, *Clementine* was the first lunar probe to carry radar since Apollo. Radar was not designed into the craft, but innovated into the mission after launch by rigging a *bistatic* system, with the transmitter in one place (the communications channel on *Clementine*) and the receiver elsewhere (the Deep Space Network antennae back on Earth). [37] By watching how *Clementine*'s radio signals bounced off the Moon, one could look for ice.

[36] *Explorer 49* went into lunar orbit six months after *Apollo 17*. Its primary mission was radio astronomy, not lunar science. The only other mission to visit the Moon between Apollo and *Clementine* was the Japanese *Hiten* in October 1991. It was primarily a technology demonstration and carried only one scientific instrument, the Munich Dust Counter.

[37] Bistatic lunar radar was performed for other reasons on *Lunar Orbiter 1*, *Luna 11* and *12* in 1966, *Explorer 35* in 1967 – 1969, *Luna 14* in 1968 and *Apollo 14*, *15* and *16* in 1971 – 1972. See "Spacecraft Studies of Planetary Surfaces Using Bistatic Radar" by Richard A. Simpson, 1993, IEEE Transactions on Geoscience & Remote Sensing, 31, 465 for the history and theory.

How can radar detect ice? Radio waves will not reflect from a uniform medium; so, radar is essentially sensitive to non-uniformities in material (such as a storm cell among the clouds, or an airplane suspended in the air). Electromagnetic radiation, such as radio waves, can be imbued with *circular polarization*, in which the direction of the electric part of the disturbance contained in the wave rotates in either a clockwise or counterclockwise direction (otherwise known as right-handed circular polarization – RCP, or left-handed – LCP, respectively). *Clementine* was built to transmit only RCP radio waves. If its RCP radio reflected once from a surface, the polarization would be reversed and the radio signal would return LCP. If the radio signal bounces twice, it tends to come back RCP. Bouncing multiple times, it could come back in a mixture of RCP and LCP. Pure ice will scatter few waves within its volume, but ice containing imbedded rocks will scatter radar more effectively. Before the wave heads outwards to the receiver, it will undergo more bounces than a wave reflecting from the top of the regolith at the Moon's surface. An RCP wave reflecting from the surface tends to come back more LCP than one reflecting from an ice/rock mix, which will be a combination of RCP and LCP. A volume of pure regolith will tend to scatter radar less strongly, and will tend to return LCP. When the angle between the radar transmitter, target and receiver is allowed to change, the ratio of RCP to LCP will change much differently for ice and regolith than it will for regolith alone. This ratio of RCP to LCP is known as the *circular polarization ratio*, or CPR.

CPR can be ambiguous in detecting ice, in that a regolith surface replete with rocks will scatter more times than one without, partially mimicking the effects of rocky ice. This has been the source of much controversy: the *Clementine* bistatic radar experiment claimed strong evidence of water in 1994, and buttressed this with other data in 2001. [38] Another radar observation with different set of angles to the target, this time with the transmitter and receiver both on Earth, produced contradictory results. This investigation finds no correlation between the CPR and the ability of lunar terrain to hold on to ice by remaining in deep freeze. The authors interpret this as meaning that there are no slabs or "lakes" of ice, but at best isolated crystals. [39]

[38] "The Clementine Bistatic Radar Experiment" by S. Nozette, C.L. Lichtenberg, P. Spudis, R. Bonner, W. Ort, E. Malaret, M. Robinson & E.M. Shoemaker, 1996, *Science*, 274, 1495; and "Integration of lunar polar remote-sensing data sets: Evidence for ice at the lunar south pole" by Stewart Nozette, Paul D. Spudis, Mark S. Robinson, D.B.J. Bussey, Chris Lichtenberg & Robert Bonner, 2001, *Journal of Geophysical Research*, 106, 23253.

[39] "No evidence for thick deposits of ice at the lunar south pole" by Donald B. Campbell, Bruce A. Campbell, Lynn M. Carter, Jean-Luc Margot & Nicholas J.S. Stacy, 2006, *Nature*, 433, 835.

Why might we expect water ice trapped in craters near the poles? The idea dates to Urey's 1952 book: some areas near the poles never see sunshine, so condensed volatiles might exist there. [40] The Moon has essentially no seasons (with the lunar rotation axis tilted only 1.6° from perpendicular to Earth's orbital plane around the Sun); so, one lunar day (a month, actually) is the nearly the same as any other. One day without sunlight implies all days are dark. In 1961 Watson, Murray and Brown showed how water vapor might stick in such cold traps for geologically long times. Water, even more than many heavier substances such as sulfur dioxide, carbon dioxide or hydrogen chloride (although not mercury), would stick longer in these polar cold traps. [41] Water forms ice easily e.g., 78 degrees C warmer than for ammonia, with almost the same mass (usually 18 atomic mass units for $H_2O$, 17 for $NH_3$), and is so common (with hydrogen, the most common element in the Universe, and oxygen, composing 45% of the mass in lunar soil), one expects water ice to form the foundation of volatiles frozen near the lunar poles.

[40] *The Planets: Their Origin & Development* by Harold C. Urey, 1952, New Haven: Yale University Press, p. 17.

[41] "On the Possible Presence of Ice on the Moon" by Kenneth Watson, Bruce Murray & Harrison Brown, 1961, *Journal of Geophysical Research*, 66, 1598, and "The Behavior of Volatiles on the Lunar Surface" by the same authors, 1961, *Journal of Geophysical Research*, 66, 3003.

    Hydrogen, being the lightest element, offers another detection scheme. When cosmic rays (here, atomic nuclei) from distant space interact with lunar soil, they often knock off or "spall" a neutron, which has the same mass as a hydrogen atom (to within 0.1%). These neutrons bounce off other nuclei in the regolith. If a neutron collides with a heavier nucleus (like oxygen), it recoils with most of its momentum, like a ball bouncing from a wall. If a neutron strikes a hydrogen atom, in contrast, it can lose much of its momentum, just like a cue ball on a pool table can hit another ball and come to an instant stop. These spallation neutrons can start with kinetic energies of millions of electron Volts (eV), whereas only 5 eV are needed to knock a hydrogen atom from a water molecule. In comparison the kinetic energy of a neutron at lunar regolith temperatures is about 0.02 eV. In between, the neutron is *epithermal*, with tens, hundreds, or even thousands of electron Volts. An epithermal neutron usually bounces out of the regolith after a few collisions *unless* there is hydrogen in the soil. Few other ways exist to absorb epithermal neutrons.

    The first epithermal neutron detector to look for hydrogen on the Moon (indeed any extraterrestrial body) was the Neutron Spectrometer on *Lunar Prospector* in 1998. It mapped epithermal neutrons over the entire lunar surface, since it entered polar orbit, like *Clementine*, and so spent a much of its 18-month mission over the poles. It detected a deficit in epithermal neutrons as large as 4.6%, concentrated in a spot at the North Pole and more spread out at the South Pole. The investigators estimated that it detected hundreds of millions of tonnes of water, but that number is model dependent, perhaps up to several billion tonnes. [42]

[42] "Fluxes of Fast and Epithermal Neutrons from Lunar Prospector: Evidence for Water Ice at the Lunar Poles" by W.C. Feldman, S. Maurice, A.B. Binder, B.L. Barraclough,

R.C. Elphic and D.J. Lawrence 1998, *Science*, 281, 1496.

In determining the amount of water detected by epithermal neutron absorption, details matter. Controversy ensued over whether *Lunar Prospector* had detected lunar hydrogen at all. For instance, fewer epithermal neutrons should accompany more thermal neutrons, not reported in the 1998 *Lunar Prospector* paper. [43] It is complicated: if the neutrons penetrate too deeply, they never exit the soil. Most neutrons sample less than 1 meter into the regolith, while some sensitivity extends 2 meters deep. This depends on the energy of the cosmic rays and their neutrons; higher energy neutrons are less affected by hydrogen. In fact above 1,000,000 electron Volts, neutrons are more sensitive to oxygen than hydrogen. Of course, elements other than hydrogen can affect neutron flux. Iron is significant, and other elements such as rare-Earths gadolinium or samarium correlate with epithermal effects. Hodges noted abnormally high $SiO_2$ content could mimic hydrogen. Nonetheless, one can detect hydrogen and use different energies to decide how deep it is in the soil. *Lunar Prospector* results indicated that potential water on the Moon likely rests a large fraction of a meter underground, covered by dry soil. Later probes imply similar results; we will deal with those soon.

[43] "Reanalysis of Lunar Prospector neutron spectrometer observations over the lunar poles" by R. Richard Hodges, Jr., 2002, *Journal of Geophysical Research*, 107, 5125; "Improved modeling of Lunar Prospector neutron spectrometer data: Implications for hydrogen deposits at the lunar poles" by David J. Lawrence, W.C. Feldman, R.C. Elphic, J.J. Hagerty, S. Maurice, G.W. McKinney & T.H. Prettyman, 2006, *Journal of Geophysical Research*, 111, E08001; "Correlation of Lunar South Polar Epithermal Neutron Maps: Lunar Exploration Neutron Detector and Lunar Prospector Neutron Spectrometer" by T.P. McClanahan, 2010, *Lunar & Planetary Science Conference*, 41, 1395.

While some researchers used epithermal neutrons and radar to study water on the Moon, other techniques developed. One evokes that used by Akhmanova and collaborators on *Luna 24* samples: absorption of light by hydration of the mineral lattice of lunar dust grains. The absorption band used for *Luna 24* occurs at wavelengths around 2.9 microns in the near infrared, caused by increasing the energy of vibration (and rotation) of the water molecule. Since these vibrations correspond in part to stretching and shrinking the bond between oxygen and one of the hydrogen atoms, a similar wavelength of light is also absorbed by the free radical, hydroxyl (OH). Water (and hydroxyl) also appears in Earth's atmosphere; so, a similar atmospheric band absorbs at many of these wavelengths (although not an identical list, since a water molecule is distorted within the mineral matrix compared to Earth's atmosphere).

Faith Vilas and collaborators investigated several absorption bands due to hydration in phyllosilicates. These are common minerals on Earth but rare on the Moon, and consist of sheets of silicates containing another metal (magnesium, iron, aluminum, manganese, potassium, calcium, lithium) and much hydroxyl, and include mica, serpentine, chlorite and clays, such as talc. These also show the 2.9-micron band, plus overtones at 1/2, 2/3 and 3/4 of this wavelength, and another at 0.7 micron due to changes in ionized iron ($Fe^{2+}$ into $Fe^{3+}$) seen in hydrated minerals. This 0.7-micron feature was used to classify phyllosilicates in the laboratory and identify them in water-bearing asteroids. [44] Vilas and collaborators published several papers showing that this absorption is seen near the lunar poles. [45] Signals of hydrated regolith seem strongest on surfaces facing the Sun, implying hydrogen implanted as energetic protons from the solar wind. The authors published the results as a refereed paper only after a decade of struggle, and only then in a relatively obscure journal unavailable online (unless one is a paid subscriber). Most lunar scientists remained unaware of this result, but not due the authors' obscurity; astronomers know Dr. Vilas as former director of the MMT Observatory, with one of the world's largest optical telescopes. Some readers dismissed the result as possibly explained by other mineral features faking phyllosilicates; the idea of hydrated regolith was too extreme.

[44] "Classification of Iron Bearing Phyllosilicates Based on Ferric and Ferrous Iron Absorption Bands in the 400-1300 nm Region" by L. Stewart, E. Cloutis, J. Bishop, M. Craig, L. Kaletzke & K. McCormack, 2006, *Lunar & Planetary Science Conference*, 37, 2185; "Phyllosilicate Absorption Features in Main-Belt and Outer-Belt Asteroid Reflectance Spectra" by Faith Vilas & Michael J. Gaffney, 2006, *Science*, 246, 790.

[45] "A Search for Phyllosilicates Near the Lunar South Pole" by E.A. Jensen, F. Vilas, D.L. Domingue, K.R. Stockstill, C.R. Coombs & L.A. McFadden, 1996, *Bulletin of American Astronomical Society: Division of Planetary Sciences*, 28, 1123; "Evidence for Phyllo-silicates near the Lunar South Pole" by Faith Vilas, E. Jensen, Deborah Domingue, L. McFadden,

Cassandra Coombs & Wendell Mendell, 1998, *Workshop on New Views of the Moon*, Houston: Lunar & Planetary Institute, p. 73; "Aqueous Alteration on the Moon" by F. Vilas, D.L. Domingue, E.A. Jensen, L.A. McFadden, C.R. Coombs & W.W. Mendell, 1999, *Lunar & Planetary Science Conference*, 30, 1343; "A newly-identified spectral reflectance signature near the lunar South pole & the South Pole-Aitken Basin" by F. Vilas, E.A. Jensen, D.L. Domingue, L.A. McFadden, C.J. Runyon & W.W. Mendell, 2008, *Earth, Planets & Space*, 60, 67.

    Vilas and collaborators used images from *Galileo*'s flyby of the Moon (on the way to Jupiter), and confirmation of suspected regolith hydration came from another spacecraft, although to the great surprise of lunar scientists. India's first lunar probe, *Chandrayaan-1*, carried the U.S.-made Moon Mineralogy Mapper ($M^3$, for short). While not specified to find hydration, that is indeed what $M^3$ saw. To confirm this surprising result, investigators called upon data from two other craft having flown by the Moon: *Deep Impact* heading to comet Tempel 1 and *Cassini* on its way to Saturn. $M^3$ was a spectrometer operating in the visual to infrared wavelength range, from 0.4 to 3.0 microns, splitting light into 260 wavelength bins or "colors." At any moment $M^3$ would view a rectangle of Moon about 40 kilometers by 0.07 kilometers and split this image into 600 individual spectra. An instant later $M^3$ would view the next 0.07 kilometer-wide strip passing beneath *Chandrayaan-1* and slowly build an image of the entire lunar surface (requiring about 300 parallel swaths) in each of the 260 colors. While the builders of $M^3$ were aware of issues concerning water, [46] the wavelengths covered actually cut off amid the 2.9-micron hydration band. The amazing result from $M^3$ was that hydration seemed ubiquitous across the lunar surface, stronger near the poles. (The 3 micron cutoff, however, made characterizing the nature of hydration, $H_2O$ versus OH, difficult.) This was so unexpected that earlier 3 microns images taken by *Deep Impact* and *Cassini* were used to confirm it: hydration is everywhere across the Moon, varying in longitude depending on the angle of the illumination of the Sun in a way which seemed to imply that the solar wind might be responsible for making it. However, $M^3$ found variation towards the poles (where the Sun barely shines) in the opposite sense: some places near the poles showed a concentration of OH (or $H_2O$) as large as about 0.07%, whereas

near the equator concentrations were typically 0.002%.  *Deep Impact* and *Cassini* supported this (without resolving the $H_2O$ versus OH issue). [47]

[46] See, for instance, http://m3.jpl.nasa.gov/Volatiles/ (retrieved before *Chandrayaan-1* mission).

[47] "Character and Spatial Distribution of OH/$H_2O$ on the Surface of the Moon Seen by $M^3$ on *Chandrayaan-1*" by C.M. Pieters et al., 2009, *Science*, 326, 568; "Detection of Adsorbed Water and Hydroxyl on the Moon" by Roger N. Clark, 2009, *Science*, 326, 562; "Temporal and Spatial Variability of Lunar Hydration As Observed by the Deep Impact Spacecraft" by Jessica M. Sunshine, Tony L. Farnham, Lori M. Feaga, Olivier Groussin, Frédéric Merlin, Ralph E. Milliken & Michael F. A'Hearn, 2009, *Science*, 326, 565.

The *Chandrayaan-1*/$M^3$ hydration result in 2009 created much excitement in the U.S and India.  Here was a detection of lunar hydration by a NASA effort that received the public endorsement of the agency, and also demonstrated scientific success for India's first probe beyond Earth, despite its demise due to overheating and power failure.  This was the first time many were aware lunar water (or at least hydroxyl) had been detected.  In fact *Chandrayaan-1* deployed a second craft, the *Moon Impact Probe*, ten months before the $M^3$ announcement, and it too found water (but this result was not revealed until the $M^3$ publication).  *MIP* carried a mass spectrometer to detect gas (CHACE; Chandra Altitudinal Composition Explorer) and found water from 45° latitude and 98 kilometers altitude down to its impact in Shackleton crater near the South Pole, the first time since the marginal detections of Apollo that any instrument had found water vapor at the Moon.   Unfortunately the gases detected are consistent with contamination by rocket exhaust. [48]

[48] *Chandrayaan-1*'s main engine and thrusters used monomethyl hydrazine and Mixed Oxides of Nitrogen (MON-3; 97% dinitrogen tetroxide): http://www.isro.org/chandrayaan1/page19.aspx. These make nitrogen, carbon dioxide and water: $4CH_3NHNH_2 + 5N_2O_4 \rightarrow 9N_2 + 4CO_2 + 12H_2O$, with masses 18, 28 and 44 amu.  CHACE saw peaks at 17-18, 28 and 44 amu, as well as 1 amu. ("The sunlit lunar atmosphere: A comprehensive study by CHACE on the Moon Impact Probe of *Chandrayaan-1*" by R. Sridharan, S.M. Ahmed, Tirtha Pratim Das, P. Sreelatha, P. Pradeepkumar, Neha Naik, Gogulapati Supriya, 2010, *Planetary & Space Science*, 58, 1567.)  ALSEP (LACE on *Apollo 17*) found little $H_2O$, $N_2$ or $CO_2$ but in this mass range mainly $^{40}Ar$ and $^{20}Ne$, which should also dominate at CHACE's altitude.  ("Molecular gas species in the lunar atmosphere" by J.H. Hoffman & R.R. Hodges, Jr,, 1975, *Moon*, 14, 159.)  The CHACE signal appears contaminated.

We should be clear: before *Chandrayaan-1* and separate from radar, neutron or infrared spectroscopy we knew that lunar water exists and came from *inside* the Moon. It is in the rocks. Despite skepticism about hydrated Apollo samples, we knew by 2008 that surprisingly large amounts of water/hydroxyl are locked in some lunar minerals. (See below.) This derives from better instrumentation and techniques to dissect contents internal to rock samples, clarifying that some minerals have prodigious amounts of water, primarily in samples from Apollo.

Previously one would vaporize a sample for analysis, ionize it with an electron beam, then run these charged particles through electromagnetic fields in a mass analyzer to measure the charge/mass ratio for the atoms, molecules and molecular fragments that result. A newer technique is SIMS or "secondary ion mass spectrometry" using an ion gun to blast tiny sub-samples from material being studied, then sending the resulting shrapnel of ions into the mass analyzer. The ion gun can be focused to a small spot that is scanned across the sample, analyzing each particle blast and constructing a map of each constituent across the sample face. Since rocks often contain a jumble of tiny mineral inclusions or grains, this is a powerful way to study each mineral separately. (Electron beams carry more charge per unit kinetic energy, so may distort the charge distribution across the sample in troublesome ways.) One can study compositional variations on scales of microns, smaller than most grains.

Erik Hauri of the Carnegie Institution improved SIMS for volatiles and with Alberto Saal at Brown University and collaborators persuaded NASA in 2007 (after years of trying) to fund turning the technique to lunar materials. Their samples were *picritic* glasses: picrite composed of olivine and pyroxene from deep in the Moon, and glasses quenched by sudden cooling at the lunar surface. The samples were in fact tiny glass spheres from *fire fountains*: lava droplets spewing into space, solidifying, then hitting the ground as beads. Expanding gas propelled these eruptions; what could that gas be, coming from deep inside the Moon?

The surprise was gas erupting from the Moon that was volcanic in composition: high in sulfur dioxide and water, even if low in carbon dioxide or monoxide.  The samples were fire fountain beads, typically a few tenth of a millimeter in diameter, from *Apollo 15*: the green glass (indicative of olivine), and *Apollo 17*'s famous orange soil (ilmenite and olivine).  By slicing the beads in half and pecking a ion microprobe every 15 microns across their diameter, Saal and collaborators showed that volatiles were concentrated on the *interior* of the beads: volatiles had leaked out (presumably when the droplets were flying through the vacuum), not leaked in (due to contamination).  The amounts were large: 115 to 576 parts per million of sulfur (presumably associated with $SO_2$) and 4 to 46 ppm of water.  There was little chlorine (0.06 – 2 ppm) but significant fluorine (4 – 40 ppm). [49] Surprisingly, there was almost no appreciable carbon dioxide, which is expected in many theoretical models. [50] Modeling how much of each gas is lost based on their concentration profile across the beads, Saal and company extrapolate that 260 – 745 ppm of water were originally present in the beads, and up to about 700 ppm sulfur.  This is a huge amount of volatiles, comparable to the amount found in basalt extruded from mid-ocean ridges on Earth. [51]

[49] "The Volatile Contents ($CO_2$, $H_2O$, F, S, Cl) of the Lunar Picritic Glasses" by A.E. Saal, E.H. Hauri, M.J. Rutherford & R.F. Cooper, 2007, *Lunar & Planetary Science Conference*, 38, 2148; "Volatile content of lunar volcanic glasses & the presence of water in the Moon's interior" by Alberto E. Saal, Erik H. Hauri, Mauro L. Cascio, James A. Van Orman, Malcolm C. Rutherford & Reid F. Cooper, 2008, *Nature*, 454, 192.

[50] "The driving mechanism of lunar pyroclastic eruptions inferred from the oxygen fugacity behavior of *Apollo 17* orange glass" by Motoaki Sato, 1979, *Lunar & Planetary Science Conference*, 10, 311.  (See page 321 especially.)

[51] "Recycled dehydrated lithosphere observed in plume-influenced mid-ocean-ridge basalts" by Jacqueline Eaby Dixon, Loretta Leist, Charles Langmuir & Jean-Guy Schilling, 2002, *Nature*, 420, 385.

Some investigators suspected that picritic glass beads might be anomalous, but hydration is also high in some crystalline lunar minerals.  Francis McCubbin of Carnegie and The University of New Mexico and his collaborators have studied apatite, which is a lattice of calcium and phosphate, with locations filled by negative ions (hydroxide, fluoride, chloride or bromide).  Apatite is well known because it happens to define level 5 on the "scratch" or Mohs scale of mineral hardness.  Using samples from *Apollo 14* and *15*, the researchers found hydroxyl at levels of 220 to 2700 parts per million, and up to 7000 ppm (or 0.7%) for lunar meteorites recovered in northwest African deserts. [52]

[52] "Nominally hydrous magmatism on the Moon" by Francis M. McCubbin, Andrew Steele, Erik H. Hauri, Hanna Nekvasil, Shigeru Yamashita & Russell J. Hemley, 2010, *Proceedings of the National Academy of Sciences*, 107, 11223.

Apatite forms at relatively low pressures, probably within about ten kilometers of the lunar surface (accounting for the lower overburden pressure of rocks in lunar gravity).  The magma melt from which apatite derives originates deeper down and are different; accounting for this the water content of the whole Moon might be much lower.  McCubbin and collaborators estimate between about 0.05 and 17 ppm water for the magma source in the interior.  This is still higher than magma water content estimated during the Apollo era, by about a factor of 100 or more.  (In contrast Saal et al. results imply 2 – 20 ppm.)  Hauri, Saal and collaborators published a later analysis based not on glass beads, but small crystals of minerals within them, not degassed, which likely preserve undiluted magmatic water content. [53] This reached even higher volatile concentrations: 615 – 1410 ppm water, 612 – 887 ppm sulfur, and 50 – 68 ppm fluorine (plus 1.5 – 3 ppm chlorine).  Correcting to the lunar mantle melt concentrations, these correspond to 79 – 409 ppm water, 193 – 352 sulfur, 7 – 26 fluorine and 0.14 – 0.83 ppm chlorine.

[53] "High Pre-Eruptive Water Contents Preserved in Lunar Melt Inclusions" by Erik H. Hauri, Thomas Weinreich, Alberto E. Saal, Malcolm C. Rutherford & James A. Van Orman, 2011, *Science Express*, 10.1126/science.1204626

Results from McCubbin et al., and Hauri, Saal and collaborators since 2007 revolutionize our understanding of the volatile content of the lunar interior (especially water).  As recently as 2006 the settled value for the lunar bulk water content was below 1 part per *billion*. [54] Most values now discussed well exceed 1 part per *million*.  This implies radical changes (see Part 2). Before these results sank in, by 2010, the attitude of most lunar scientists regarding water inside the Moon was "extraordinary claims demand extraordinary evidence." Was the idea of water in the Moon so extraordinary? No. Beyond the Sun every major Solar System body shows the presence of water, as do many lesser ones, including all comets and many asteroids. Mercury, with daytime temperatures near 430°C but in many ways similar to the Moon, hordes at least a few billion tonnes of water ice at its poles, and shows pyroclastic features indicating magma volatiles composing several tenths of a percent, like terrestrial volcanoes. [55] Earth, which also would be baked by a Moon-forming impact, is the "water planet." From whence came this water? [56] The Big Whack seems to have desiccated the Moon, but to what extent? Are there loopholes? Besides, this is theory, and a relatively young one (compared to, for instance, the five-decade wait in accepting continental drift). Before 2007 evidence did not rule out lunar water to the extent commonly held, and left open the prospect of gas, even water vapor, leaking recently from the Moon. This seems even more likely now that we know water helped drive lunar fire fountain eruptions. Much remains unknown about the Moon. This has long fascinated me, since 2006 been a major research effort of mine, and will be a focus of Part 3.

[54] "Earth-Moon System, Planetary Science & Lessons Learned" by S. Ross Taylor, Carle M. Pieters & Glenn J. MacPherson, 2006, in *New Views of the Moon*, edited by Bradley L. Jolliff, Mark A. Wieczorek, Charles K. Shearer & Clive R. Neal (Chantilly, VA: Mineralogical Society of America), pp. 657–704, summarizing on page 663: "The Moon is dry, with less than one ppb water, except for some possible amounts trapped in permanently shadowed craters …"

[55] "Explosive volcanic eruptions on Mercury: Eruption conditions, magma volatile content & implications for interior volatile abundances" by Laura Kerber, James W. Head, Sean C. Solomon, Scott L. Murchie, David T. Blewett & Lionel Wilson, 2009, *Earth & Planetary Sciences Letters*, 285, 263; the mass of Mercury's polar ice is uncertain, but a lower limit is estimated ("External Sources of Water for Mercury's Putative Ice Deposits" by Julianne I. Moses, Katherine Rawlins, Kevin Zahnle & Luke Dones, 1998, *Icarus*, 137, 197).

[56] The origin of Earth's water is unknown. It could pre-date the Giant Impact and/or be cometary, meteoritic, chemical or even biological in origin, or all of these. We discuss this further below.

The recent lunar water data have been challenged, for which chlorine is crucial. Hydrogen and chlorine combine strongly, and hydrochloric acid (HCl) reacts strongly in magma melts. Chlorine comes in two stable isotopes, $^{35}$Cl and $^{37}$Cl (with 18 and 20 neutrons, respectively), which on Earth everywhere are split in a 76.8:24.2 abundance ratio, nearly without exception. Being lighter, $^{35}$Cl vaporizes faster, but $^{37}$Cl reacts more easily with H. Together these factors cancel in terms of chlorine isotopes remaining in a melt or vapor, if hydrogen is present in greater abundance. Absent H, however, this balance is broken, and $^{37}$Cl/$^{35}$Cl ratios can vary. This is seen on the Moon: some minerals have the same $^{37}$Cl/$^{35}$Cl ratio as Earth's, but others have up to 2.4% more $^{37}$Cl. [57] This may indicate that hydrogen is rarer than chlorine, despite hydration results. An alternative is that the Moon is non-uniform, either in its initial $^{37}$Cl/$^{35}$Cl ratio or hydrogen concentration. In this paper non-uniformity is dismissed due to homogenization in the magma ocean. There is no discussion of *some* magmas being hydrated.

[57] "The Chlorine Isotope Composition of the Moon & Implications for an Anhydrous Mantle" by Z.D. Sharp, C.K. Shearer, K.D. McKeegan, J.D. Barnes & Y.Q. Wang, 2010, *Science*, 329, 1050.

The weight of the evidence indicates that there are at least two and probably three sources of hydration on the Moon. [58] There is water (or hydroxyl) imbedded in some minerals (Saal, Hauri; McCubbin and collaborators). There is hydration that appears on the surface, as evidenced by infrared and optical absorption, and may be due to solar wind hydrogen reacting with regolithic oxygen (Vilas; Pieters and collaborators). Asteroids and comets, which contain water, undoubtedly strike the Moon and presumably add water vapor to its atmosphere, temporarily. If those water molecules (or any other) enter the ultra-cold craters permanently shadowed near the poles, they will stick. Water could also come from other sources.

[58] Suggestions include water escaping Earth's atmosphere, reaching the Moon via the magnetotail ("The Source of Water Molecules in the Vicinity of the Moon" by T. Földi & Sz. Bèrczi, 2001, *Lunar & Planetary Science Conference*, 32, 1148), and water brought occasionally into the Solar System by interstellar giant molecular clouds. See Part 2. Interplanetary dust can

also be hydrated, and strikes the Moon. ("The Poles of the Moon" by Paul G. Lucey, 2009, *Elements*, 5, 41.)

In 2006 water's presence anywhere on the Moon was in doubt, at least in some scientists' minds. (I heard it said at the time that some would not believe lunar water existed until they drank it from a glass.) In April that year NASA announced a plan to probe water on the Moon almost as directly as pouring a glass. *LCROSS* (*Lunar CRater Observation and Sensing Satellite*) would slam into a permanently shadowed lunar crater and dig out tons of regolith, and hopefully water. Relatively inexpensive because it rode along with *LRO* on the same rocket, *LCROSS* in part *was* the rocket, or at least its Centaur upper stage. The Centaur was prepared to minimize contamination, emptied of fuel, and guided to its doom by the Shepherding Spacecraft (*S-S/C* or *SSc*), a modified, detachable ring joining the Centaur to the payload (*LRO*). Hitting the Moon with 2.3 tonnes at 2.5 kilometers per second, the Centaur blasted a crater about 25 meters in diameter and 4 meters deep, on 2009 October 9. *LRO* flew overhead, examining the debris. The *S-S/C*, with instruments observing the impact, followed the Centaur to a nearby impact four minutes later.

The results from *LCROSS* were surprising in several ways. The impact site was chosen for maximal hydrogen signal in neutron absorption, and maximum possible visibility from Earth (consistent with being deep in a permanently shadowed depression). Observations from Earth largely missed the event, [59] since the densest debris concentration remained hidden low behind a two-kilometer high massif at the edge of South Pole-Aitken basin. Signals seen by *S-S/C* were faint, but informative. The *S-S/C* carried five cameras covering the optical to mid-infrared, three ultraviolet to near-infrared spectrometers, and an optical photometer. Several of these detected faint emission about 0.3 seconds after impact which brightened as an increasing plume of ejecta rose to intercept the sunlight. About 10,000 tonnes were heated, typically to 1000 K, and about 5000 tonnes were thrown into the sunlight. [60] Emission lines from water were strong, indicating that about 5.6% of the mass was water. [61]

[59] Resonance line emission from two grams of sodium in the ejecta was observed from Earth. ("Observations of the lunar impact plume from the LCROSS event" by R. M. Killen, A. E. Potter, D. M. Hurley, C. Plymate & S. Naidu, 2010, *Geophysical Research Letters*, 37, L23201.

[60] "The *LCROSS* Cratering Experiment" by Peter H. Schultz, Brendan Hermalyn, Anthony Colaprete, Kimberly Ennico, Mark Shirley & William S. Marshall, 2010, *Science*, 330, 468

[61] "Detection of Water in the *LCROSS* Ejecta Plume" by Anthony Colaprete et al., 2010, *Science*, 330, 463.

    *LCROSS* is fascinating not just due to water on the Moon, less controversial by 2009, but other substances: 5.7% carbon monoxide, 1.4% molecular hydrogen, 1.6% calcium, 1.2% mercury, 0.4% magnesium. Sulfur is detected as hydrogen sulfide ($H_2S$) and $SO_2$, at levels 1/6th and 1/30th of water, respectively. Nitrogen is seen within ammonia ($NH_3$), at 1/16th water's abundance. Trace amounts (less than 1/30th of water) are detected for ethane ($C_2H_4$), $CO_2$, methanol ($CH_3OH$), methane ($CH_4$) and OH. [62] Volatiles compose at least one-tenth of the soil mass. The poles differ radically from any part of the Moon we have visited or sampled.

[62] "*LRO*-LAMP Observations of the *LCROSS* Impact Plume" by G. Randall Gladstone et al., 2010, *Science*, 330, 472; "Detection of Water in the *LCROSS* Ejecta Plume" by Anthony Colaprete et al., *Science*, 330, 463. Some on this collaboration expect some detected abundances to decline.

    *LCROSS* liberated large amounts of water, not by the glass but by several barrels full. Still, it is not water you would want to drink without some processing; it is tainted with chemical and even isotopic poisons (more about isotopes in Part 2). It is more of a carbonated soup stocked with metals and organics. All of these "contaminants" are scientifically interesting and potentially useful as resources. It is not time yet to send in a backhoe and start mining, however. There are many complicated processes that we should first understand (more about this in Part 2, also). NASA and other agencies are rapidly pursuing sending rovers and other probes into these permanently shadowed regions (PSRs). As compelling as the results from *LCROSS* may be, we should not yet jump to presuming that we now have a basic understanding of all of the important processes; there are signs of unanticipated and fundamental factors beyond the simple story. Having denied the existence of lunar

water for so long, there is a tendency to make the sudden transition to presuming that we now understand the basic story and can narrow our vision to confirming our predictions. Not so fast.

**FIGURES:**

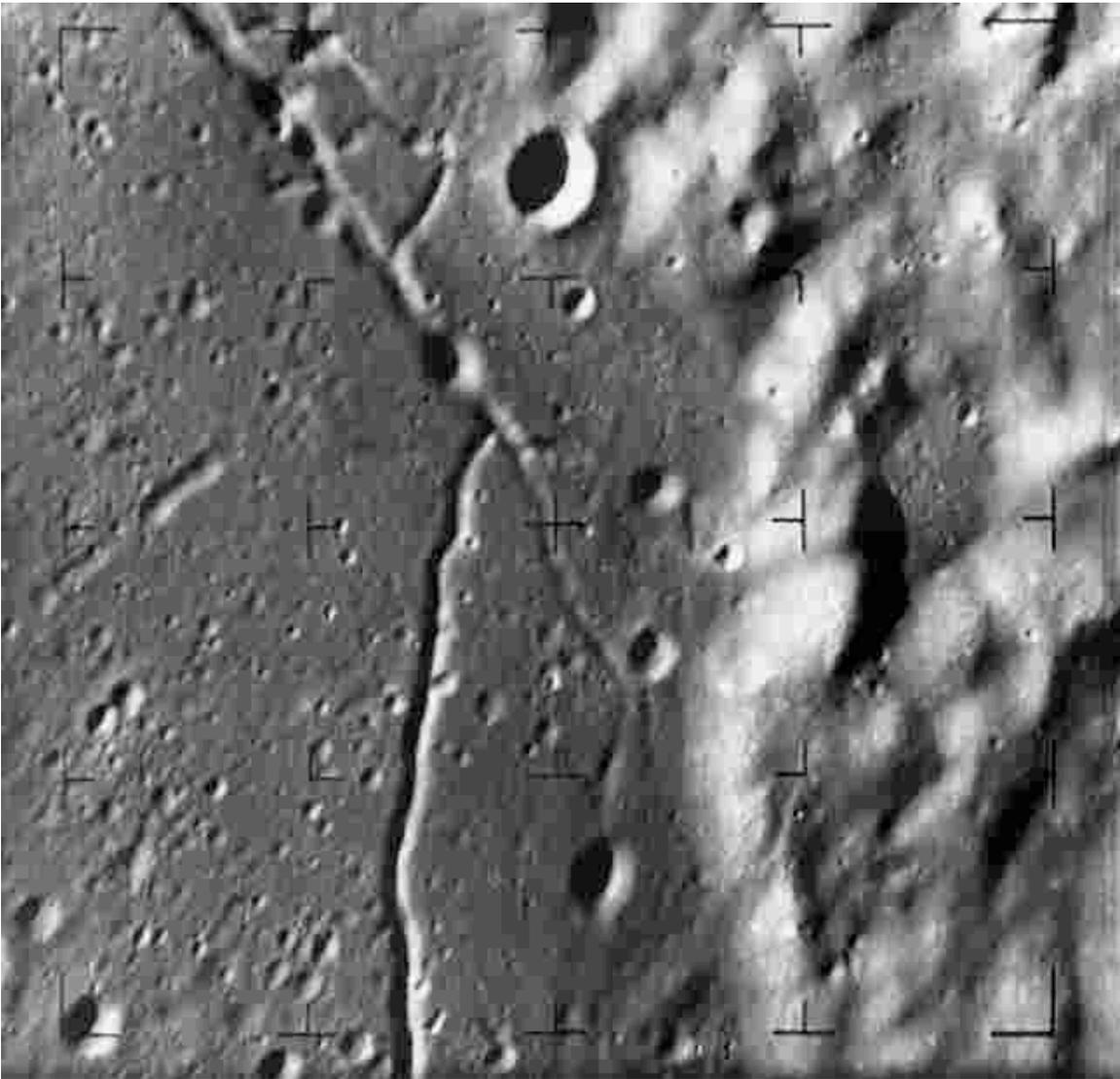

**Figure 1:** a) *Ranger 9* photograph B-074, showing a 7.3-kilometer wide portion of western Ptolemaeus crater, just north of where *Ranger 9* impacted 72 seconds later in the crater Alphonsus. Note the straight and arcuate rilles. See *Ranger IX Photographs of the Moon* by Gerard P. Kuiper, R.L. Heacock, E.M. Shoemaker, H.C. Urey & E.A. Whitaker, 1965 December 15, NASA Publication SP-112 for more details.

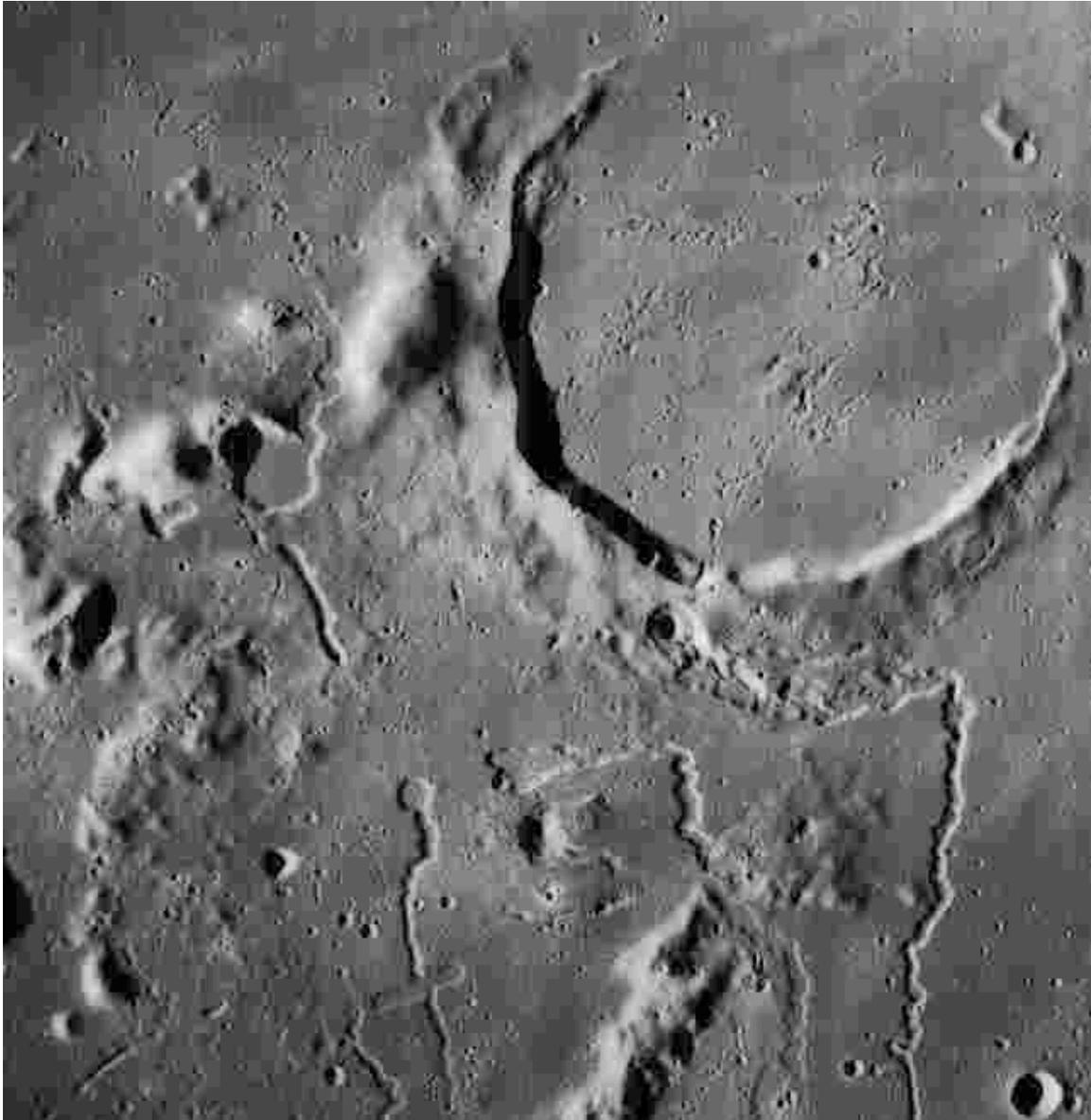

**Figure 1:** b) *Lunar Orbiter 5* photo V-189-Med showing crater Prinz (upper right), Montes Harbinger (mountains, right and bottom), and Rimae Prinz (valleys, right and bottom). Urey was heavily influenced by Lunar Orbiter images such as this showing sinuous rilles. See *Guide to Lunar Orbiter Photographs* by Thomas P. Hansen, 1970, NASA Publication SP-242.

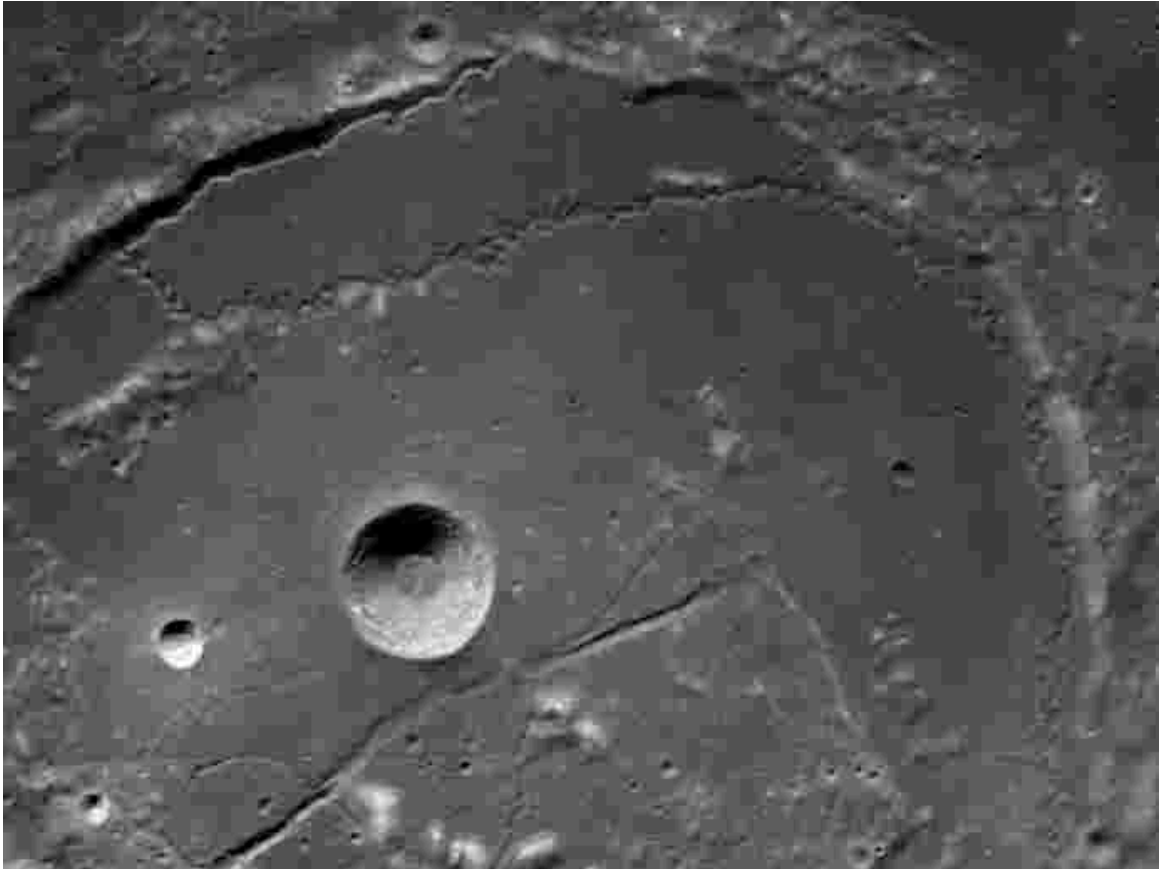
**Figure 2:** a) two views of the sinuous rille in the crater Posidonius, as seen by Lunar Reconnaissance Orbiter Camera: a) an 80-kilometer wide view (portion of mosaic of three Wide Angle Camera images) of a portion of Posidonius, showing rille Rimae Posidonius arising in the lower left corner, deepening as it meanders to the upper right and reversing itself to flow along the upper crater wall until spilling out through a gap into the mare area of Lacus Somniorum.

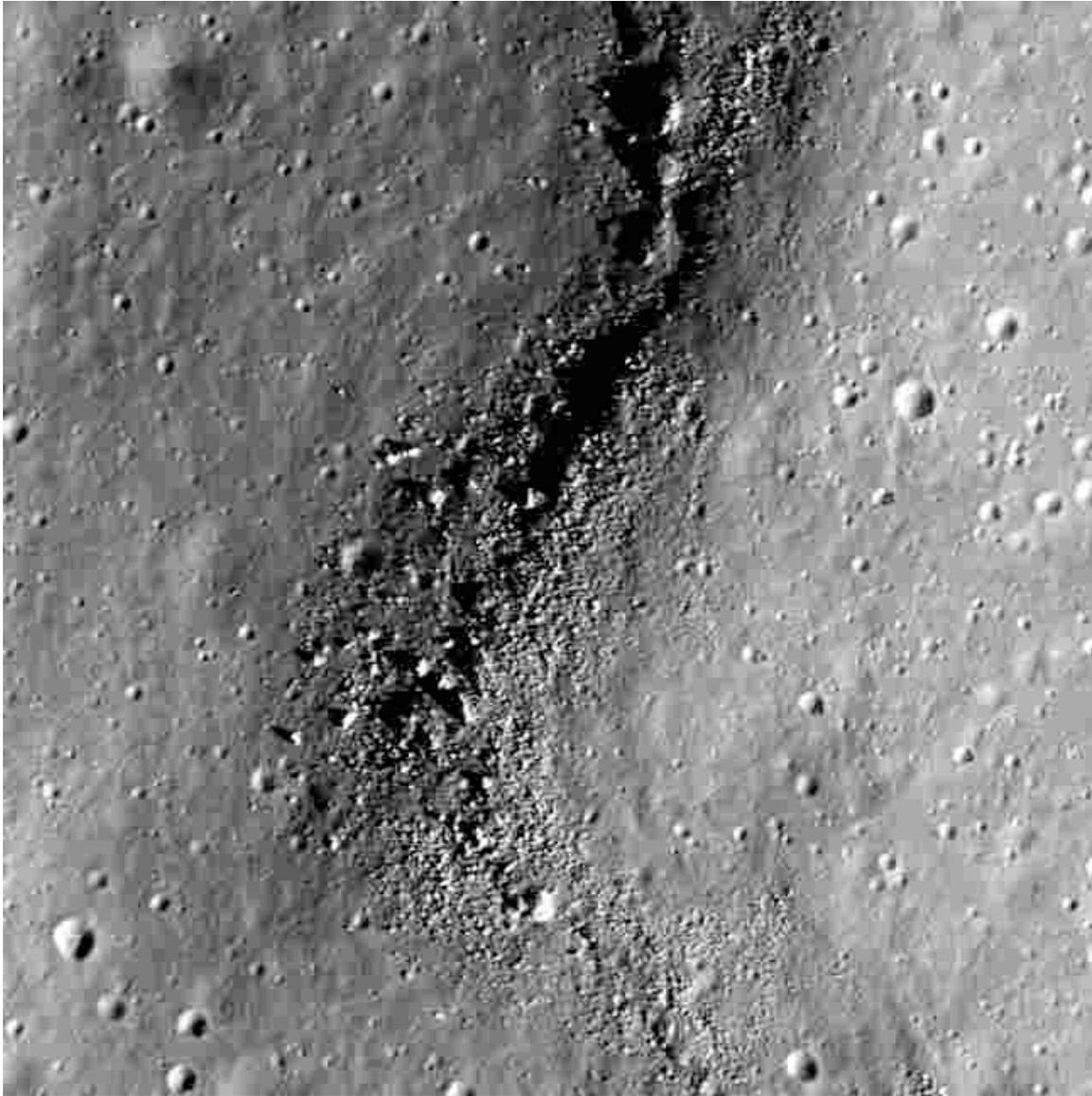

**Figure 2:** b) a 500-meter wide portion of Narrow Angle Camera image M113771795RE (zooming into the location in Figure 2a above and right of center where the rille encounters the rightmost portion of a small, horizontal ridge) showing the rille's marginal wall, indicated by mass wasting talus. Note the surface texture and cratering density is similar inside and outside the rille, supporting the idea that some rilles are lava erosional features, not collapsed lava tubes.

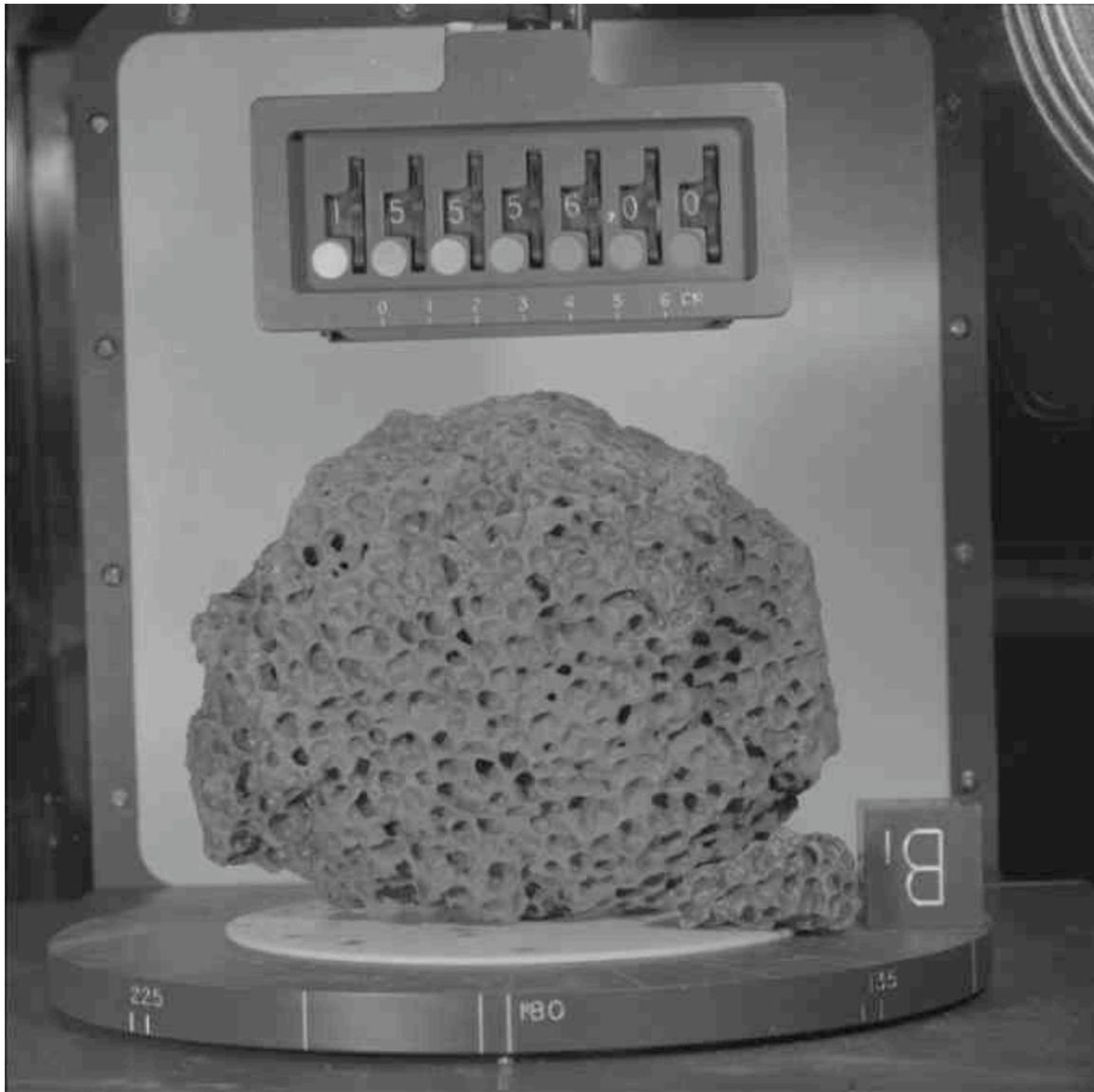

**Figure 3:** Vesicular mare basalts, in the laboratory and in situ: a) *Apollo 15* sample 15556, with a mass of 1.5 kilogram, was collected near the rim of Hadley Rille and is 3.4 billion years old. Hadley Rille is sinuous and displays layered basalt exposed in the wall of the rille.

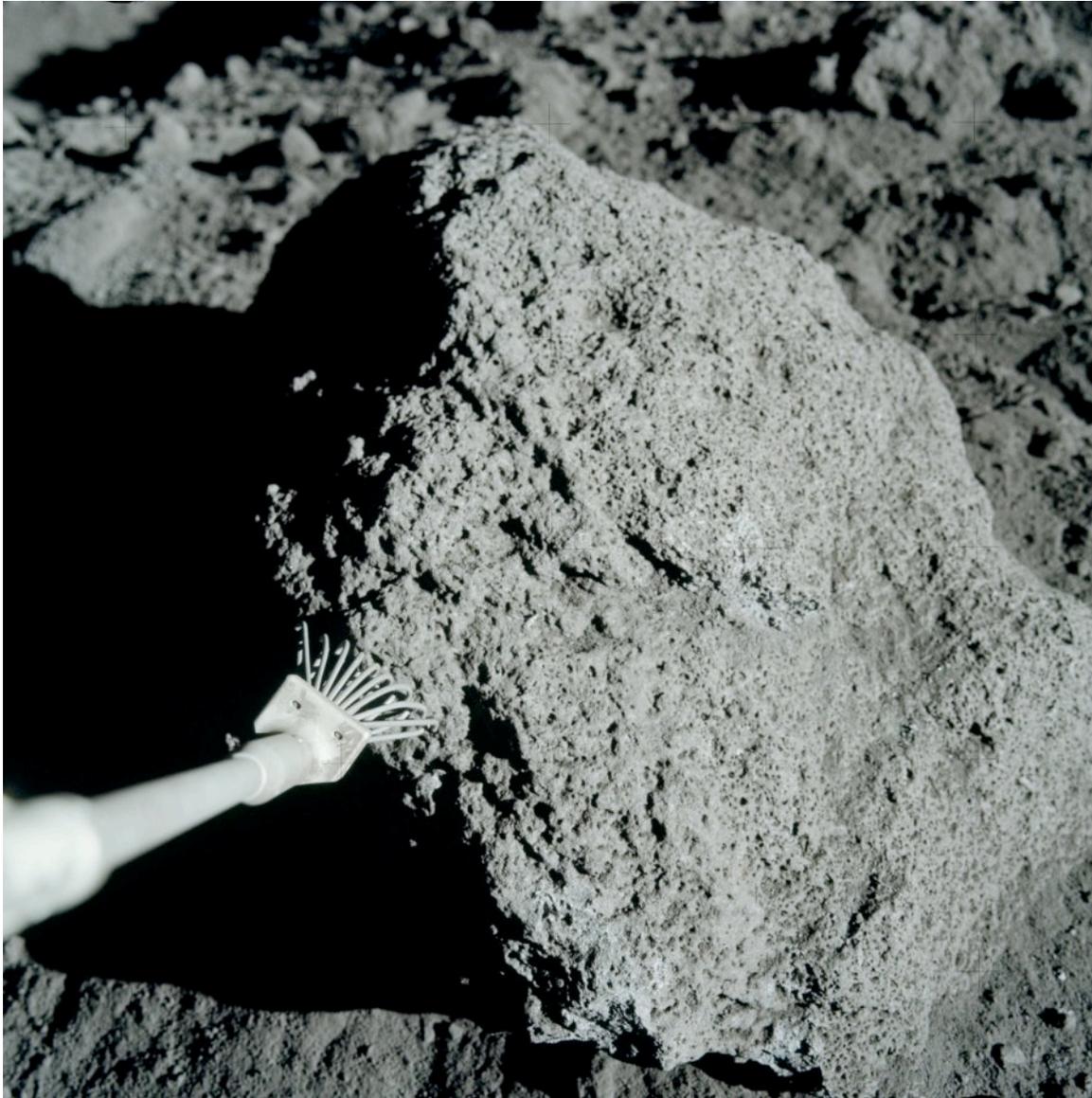
**Figure 3:** b) a vesicular basalt boulder at EVA Station 1 in the Taurus-Littrow Valley visited by *Apollo 17* (photograph AS17-134-20403). Astronaut Gene Cernan found this rock curious because "the vesicularity changes from a hummocky vesicularity to a very fine vesicular" along a boundary seen here running near the center of the rock roughly vertically (upper left to lower right) indicating an inhomogeneous change of phase. Cernan used sample tongs at left to set the camera focus distance.